\documentclass[usegraphicx]{mn2e}
 \usepackage{subfigure}
\usepackage{amsmath}
\usepackage{amssymb}
\usepackage{epstopdf}
\usepackage{epsfig}

\newcommand {\apgt} {\ {\raise-.5ex\hbox{$\buildrel>\over\sim$}}\ }
\newcommand {\aplt} {\ {\raise-.5ex\hbox{$\buildrel<\over\sim$}}\ }

\title[Broad absorption features in wind-dominated ULXs?]
{Broad absorption features in wind-dominated ULXs?}

\author[M.Middleton et al.]
{Matthew J. Middleton$^{1}$,  Dominic J. Walton$^{2}$, Timothy P. Roberts$^{3}$
 and Lucy Heil$^{1}$\\
1. Astronomical Institute Anton Pannekoek, Science Park 904,1098 XH, Amsterdam, Netherlands\\
2. Astronomy Department, California Institute of Technology, 1200 East California Boulevard, Pasadena, California 91125, USA\\
3. Department of Physics, University of Durham, Durham DH1 3LE, UK
}

\pagerange{\pageref{firstpage}--\pageref{lastpage}} \pubyear{2011}
\long\def\symbolfootnote[#1]#2{\begingroup\def\thefootnote{\fnsymbol{footnote}}\footnote[#1]{#2}\endgroup} 
\def\ga{\mathrel{\hbox{\rlap{\hbox{\lower4pt\hbox{$\sim$}}}{\raise2pt\hbox{$>$}}
}}}

\begin{document}

\topmargin = -0.5cm

\maketitle

\label{firstpage}

\begin{abstract}

The luminosities of ultraluminous
X-ray sources (ULXs) require an exotic solution with either
super-critical accretion modes onto stellar mass black holes or
sub-critical accretion onto intermediate mass black holes (IMBHs)
being invoked. Discriminating between the two is non-trivial due to
the present lack of a direct mass measurement. A key expectation of
the super-critical mode of accretion is the presence of powerful
radiatively-driven winds.  Here we
analyse {\it XMM-Newton} data from NGC~5408~X-1 and NGC~6946~X-1 and
find that strong soft residuals present in the X-ray spectra can be
reconciled with broadened, blue-shifted absorption by
a partially ionised, optically thin phase of this wind. We derive initial values for the physical parameters of 
the wind; we also discuss other possible origins
for the observed features.

\end{abstract}

\section{Introduction} 

Recent studies of ultraluminous X-ray sources (ULXs: Roberts 2007;
Feng \& Soria 2011) in nearby galaxies have shown that up to
$\sim$3$\times$10$^{39}$erg s$^{-1}$, their luminosities can be 
robustly associated with accretion around the Eddington limit onto
stellar mass ($<$ 100 M$_{\odot}$) black holes (e.g. Middleton et al. 2012,
2013). Those with luminosities
reaching $\sim$3$\times$10$^{40}$ erg s$^{-1}$ are harder to explain
yet seem to show a trend of spectral shape with variability (Sutton, Roberts
\& Middleton 2013). Their behaviour appears strikingly dissimilar the expected behaviour of 
sub-Eddington Galactic black hole binaries scaled up into the 
intermediate-mass black hole regime (IMBHs; Colbert \& Mushotzky 1999, with masses $>$100 
M$_{\odot}$), implying they cannot {\it all} be IMBHs unless the characteristic 
properties of accretion alter dramatically beyond the expectations of
simple mass scaling.

The behaviour of most bright (3-30$\times$10$^{39}$ erg s$^{-1}$) ULXs
can plausibly be explained by stellar mass black holes 
undergoing accretion in the `super-critical' regime (Shakura \&
Sunyaev 1973). In this situation, the inflow changes at the radius
where the accretion rate is locally Eddington. This drives an increase
in scale height of the disc and optically thick winds to be
radiatively launched from the surface (Poutanen et al. 2007; Dotan \&
Shaviv 2011; Ohsuga \& Mineshige 2011) which are likely to be
stratified due to the Rayleigh-Taylor or radiative-hydrodynamic
instabilities (Takeuchi et al. 2013). The presence of such winds are
predicted to affect the spectrum (Poutanen et al. 2007), observed
luminosity (King 2009) and variability (Middleton et al. 2011)
depending on inclination angle to the observer's line-of-sight and
changing mass accretion rate (which we will describe fully in a forthcoming
model).

Observations show that narrow atomic features in emission or absorption (namely Fe K$_{\alpha}$) must be intrinsically weak or simply not present in
two of the hardest (and brightest) ULXs (Walton et al. 2012;
2013). However, in the super-critical wind model, these `hard ultraluminous' ULXs are those that we view close to face-on (Sutton et al. 2013). The `soft ultra luminous' ULXs that display far higher levels of fractional variability, are thought to be viewed such that the
line-of-sight to the hottest inner regions are obscured by the
optically thick phase of the clumpy wind (i.e. scattered out of the
line-of-sight: Middleton et al. 2011; Middleton et al. in prep). It is therefore these `soft ultraluminous' sources where we should expect to see atomic absorption features, the expected signatures of
such outflows. 

Interestingly, it has 
long been known that soft spectrum ULXs can show strong residuals to their 
best fitting continuum models (e.g. Stobbart, Roberts \& Wilms 2006 and 
references therein), which have traditionally been associated with
emission from collisionally heated gas in the host Galaxy. However,
the intrinsic luminosities in such components are very large
(approaching $\sim$1$\times$10$^{39}$ erg s$^{-1}$)
and such residuals are seen in ULXs without large amounts of host
galactic diffuse emission (Bachetti et al. 2013) leading some to question
this identification (Middleton et al. 2011).

Here we investigate whether absorption in the wind can account for the spectral residuals in this subset of ULXs by
studying two of the brightest members of the population. 


\section{Wind-dominated ULXs}

The most variable ULXs detected are NGC~5408~X-1 (Middleton et
al. 2011; Heil, Vaughan \& Roberts 2009; Pasham \& Strohmayer
2012) and
NGC~6946~X-1 (Rao et al. 2010), both with energy integrated fractional excess
variances over intra-observation timescales of
$>$20\% (and up to $\sim$60\%). Both sources are also known to harbour quasi-periodic
oscillations (QPOs); although determining the mass based on
association to those QPOs seen in BHBs (e.g. Strohmayer \& Mushotzky 2009)
has proven non-trivial (Middleton et al. 2011; Pasham \& Strohmayer
2012).

The evolution of NGC 5408 X-1 over 6  deep {\it XMM-Newton}
observations (spanning several years) has been well studied by Pasham
\& Strohmayer (2012) who found very little change in the X-ray spectra
(although significant changes were found in the variability
power spectra). The X-ray spectra show a strong soft thermal component,
peaking below 1~keV with a tail of emission rolling over at 3-4~keV
(such a break appears almost ubiquitously in high quality {\it XMM-Newton} ULX data: Gladstone et
al. 2009 now confirmed by NuSTAR: Bachetti et al. 2013). In the super-critical framework the soft emission
is from the wind (and photosphere), with the hard emission originating
in a heavily distorted inner accretion disc (Poutanen et al. 2007). In
addition to models describing the continuum, past studies have
highlighted the need for an extra component to account for spectral residuals
below 2~keV (e.g. Strohmayer \& Mushotzky 2009). Should these be associated with a collisionally excited
plasma (emitting as a Brehmstrahlung spectrum with overlaid emission
lines) then the integrated luminosity is found to approach $\sim$1$\times$10$^{39}$ erg s$^{-1}$. This is considerably brighter than predicted for the star-formation-related diffuse emission of the {\it entire} galaxy, inferred to be $\sim$3$\times$10$^{37}$erg s$^{-1}$ (see Dale et al. 2005; Calzetti et al. 2007; Mineo et al. 2012). 


\begin{table}
\begin{center}
\begin{minipage}{80mm}
\caption{ULX observational information}
\begin{tabular}{l|c|c|c|c|c|c}
  \hline

ULX & OBSID & obs. date & good time \\
&&&(ks) \\
   \hline
NGC 5408 X-1   & 0302900101  & 2006-01-13 & 85.4  \\
  &  0500750101 & 2008-01-13 & 28.6 \\
   &  0653380201 & 2010-07-17 & 71.8 \\
   &  0653380301 & 2010-07-19 & 88.2  \\
   &  0653380401 & 2011-01-26 & 73.4  \\
   &  0653380501 & 2011-01-28 & 69.2  \\  
NGC 6946 X-1      & 0691570101  & 2012-10-21 & 81.1  \\
\hline
\end{tabular}
Notes: {\it XMM-Newton} observation identifier, date of exposure,
PN spectral exposure (after accounting for background flares and deadtime) for each observation studied.
\end{minipage} 

\end{center}
\end{table}

Middleton et al. (2011) proposed that the large amplitudes of rms
and trend of fractional variability with energy seen in NGC~5408~X-1
could
both be explained by optically thick, clumpy material in the wind obscuring sight lines to the
inner disc (e.g. Takeuchi et al. 2013). In this scenario both the spectra and variability can be explained by super-critical accretion and are heavily influenced by the wind. NGC~6946~X-1 is not as well-studied as NGC~5408~X-1, however, Rao et
al. (2010) have shown it to have both a similar X-ray spectrum and
similar (in fact larger) amplitudes of rms variability on the same
timescales. Given the close association of variability and spectral
behaviour in accreting sources (e.g. Mu{\~n}oz-Daria et al. 2011)
it is quite reasonable to assume both sources could be `wind-dominated' ULXs. 


\begin{table*}
\begin{center}
\begin{minipage}{180mm}
\caption{{\sc xspec} model best-fit parameters }
\begin{tabular}{l|c|c|c|c|c|c|c|c}
  \hline\hline
\multicolumn{9}{c}{\sc tbabs*(diskbb+nthcomp)}\\
\hline
ULX & $N_{\rm H}$ & \multicolumn{2}{c} {$kT_{\rm d}$} & \multicolumn{2}{c} {$\Gamma$} & \multicolumn{2}{c} {$kT_{\rm e}$} & $\chi^2$\\
   \hline
5408 X-1  &  0.092$\pm$0.003 & \multicolumn{2}{c}{0.193 $\pm$ 0.003} & \multicolumn{2}{c}{1.65$^{+0.12}_{-0.10}$} & \multicolumn{2}{c}{2.25 $\pm$ 0.04} &  2522/1663 \\
6946 X-1  &  0.304$^{+0.021}_{-0.020}$  &  \multicolumn{2}{c} {0.232$^{+0.015}_{-0.014}$} & \multicolumn{2}{c} {1.96 $\pm$ 0.11} & \multicolumn{2}{c}{1.63$^{+0.31}_{-0.20}$}  &  944/835 \\
\hline\hline

\multicolumn{9}{c}{\sc tbabs*tbvarabs*(apec+diskbb+nthcomp)}\\
\hline
ULX & $N_{\rm H}$ & $kT_{\rm apec}$ & $kT_{\rm d}$ & \multicolumn{2}{c} {$\Gamma$} & \multicolumn{2}{c} {$kT_{\rm e}$} & $\chi^2$\\
\hline

5408 X-1  & 0.052 $^{+0.009}_{-0.008}$ & 1.01 $\pm$ 0.03 & 0.152 $\pm$0.008 & \multicolumn{2}{c} {2.36 $\pm$ 0.06} &\multicolumn{2}{c} {2.09$^{+0.48}_{-0.29}$} & 1860/1661\\

5408 X-1  & 0.049 $\pm$ 0.004 & 1.02 $\pm$ 0.02 & 0.156 $\pm$ 0.004  & \multicolumn{2}{c} {2.36 $\pm$ 0.03} &\multicolumn{2}{c} {2.03$^{+0.21}_{-0.16}$} & 1894/1661\\ 


6946 X-1  & 0.028 $^{+0.023}_{-0.021}$ & 1.14$^{+0.07}_{-0.11}$ & 0.237$\pm$0.024  & \multicolumn{2}{c} {1.98 $^{+0.11}_{-0.12}$} &\multicolumn{2}{c} {1.76$^{+0.41}_{-0.25}$} & 788/833\\

6946 X-1  & 0.039 $^{+0.028}_{-0.025}$ & 1.14$^{+0.07}_{-0.09}$ & 0.225$^{+0.027}_{-0.025}$  & \multicolumn{2}{c} {2.00 $^{+0.11}_{-0.12}$} &\multicolumn{2}{c} {1.81$^{+0.45}_{-0.27}$} & 785/833\\

\hline\hline
\multicolumn{9}{c}{\sc tbabs*tbvarabs*xstar*(diskbb+nthcomp)}\\
\hline
ULX & $N_{\rm H}$ & wind $N_{\rm H}$ & $log \xi$ & $z$ & $kT_{d}$ & $\Gamma$ & $kT_e$ & $\chi^2$\\
\hline

 5408 X-1  &  0.028 $\pm$ 0.005  & 19.8$^{+23.9}_{-5.8}$ & 3.44$^{+0.14}_{-0.10}$ & -0.12 $\pm$ 0.01 & 0.208 $\pm$ 0.007 & 2.29 $\pm$ 0.08 & 1.76$^{+0.36}_{-0.23}$ & 1997/1660 \\

5408 X-1  &  0.028 $\pm$ 0.003  & 6.4$^{+8.3}_{-1.5}$ & 3.18$^{+0.05}_{-0.07}$ & -0.12 $\pm$ 0.01 & 0.206 $\pm$ 0.004 & 2.29 $^{+0.04}_{-0.05}$ & 1.73$^{+0.16}_{-0.13}$ & 1985/1660 \\

6946 X-1  &  0.115 $^{+0.014}_{-0.022}$  & 100.0 ($>$ 55.2) & 3.69$^{+0.02}_{-0.11}$ & -0.13 $^{+0.01}_{-0.02}$ & 0.237 $^{+0.019}_{-0.009}$ 
& 2.05 $^{+0.12}_{-0.14}$ & 1.95$^{+0.57}_{-0.40}$ & 863/832 \\

6946 X-1  &  0.132 $^{+0.034}_{-0.021}$  & 41.9$^{+4.2}_{-1.11}$ & 3.52$^{+0.17}_{-0.06}$ & -0.13 $\pm$ 0.01 & 0.218 $^{+0.015}_{-0.021}$ 
& 2.12 $^{+0.16}_{-0.12}$ & 2.14$^{+2.73}_{-0.46}$ & 838/832 \\

 \hline\hline

\end{tabular}
Notes: Units of $N_{\rm H}$ are 10$^{22}$ cm$^{-2}$ and temperatures (seed photons: $kT_{\rm d}$ and electron plasma: $kT_{\rm e}$) are quoted in keV.
{\it Top:} Best-fit model parameters (and 90\% errors) for the
continuum model. {\it Middle:} model parameters when including thermal
plasma emission with variable abundances (0.5~Z$_{\odot}$  top and Z$_{\odot}$ 
bottom).{\it Bottom:} when including absorption in a wind with a velocity
dispersion of 0.1c, outflow velocity of $z\times$c and 0.5~Z$_{\odot}$ and Z$_{\odot}$  (top and bottom sets of values in
each case).
\end{minipage} 
\vspace{-0.3cm}
\end{center}
\end{table*}

\section{Spectral modelling}

We re-processed the  6 {\it XMM-Newton} (EPIC MOS and PN) archival datasets of
NGC~5408~X-1 (acquired from
HEASARC (http://heasarc.gsfc.nasa.gov/) and
the latest (proprietary) observation of NGC~6946~X-1 using {\sc
  sas 12} and up-to-date current calibration files (see Table 1 for details).  We followed
standard extraction procedures and filtered for soft proton flaring
in the full-field, high energy (10-15~keV) background, extracting
good time intervals (GTI).  As the PN has the highest response to
these events we used the PN GTI as input in extracting spectra, using
{\sc xselect}, confirming that this removes all flares in the
equivalent MOS lightcurves. We selected $\ge$30 arcsec radius source and
background regions with the latter selected to be on the same chip,
free from sources and away from the readout direction of bright
sources. The OBSIDs, observation dates and useful exposures are given
in Table 1. Given that the spectra of NGC~5408~X-1 do not change by a
large amount at soft energies where the residuals are seen, we combined
the MOS (1 \& 2 separately) and PN datasets using {\sc
  addascaspec}. However, as the spectrum of NGC 6946 X-1  
changes markedly compared to earlier, shorter observations, we
focus our analysis only on the latest, deep observation.
  
We proceeded to initially model the continuum for each source in order to
identify and study the residuals. To this end we used a suitable convolved model 
in {\sc xspec} (Arnaud 1996) representative of emission from the outer wind
({\sc diskbb}) and emission from the inner hot disc ({\sc nthcomp}:
Zycki, Done \& Smith 1999). The Compton component gives us the
additional freedom to account for any putative Compton down-scattering
in the wind (Titarchuk \& Shrader 2005) which may act to distort the
hot disc profile, extending it to lower energies (Middleton et al. in
prep). We set the input seed photon temperature to be that of the
lower temperature wind to ensure correct energy balance in the event
of scattering and prevent further rollover at energies out of the
bandpass (consistent with this model of accretion:
Poutanen et al. 2007, and recently confirmed in {\it NuSTAR} observations by
Bachetti et al. 2013). We included absorption by neutral ISM material ({\sc tbabs}) with
appropriate abundances (Wilms et al. 2000) and lower limit set to the Galactic
line-of-sight column (NGC~5408: 6.0$\times$10$^{20}$ cm$^{-2}$ and
NGC~6946: 2.1$\times$10$^{21}$ cm$^{-2}$; Dickey \& Lockman 1990). As is standard
practice we also included a constant component to account for
differences between detector responses (usually $<$10\% from unity
except in the case of NGC~6946~X-1 MOS1 where a dead column across the
source affected the spectrum).


The results of the spectral fitting are given in Table 2 and plotted
in Figure 1. We confirm the presence of strong residuals in both
NGC~5408~X-1 and NGC~6946~X-1 MOS and PN spectra. Following previous
analyses of such residuals (e.g. Stobbart et al. 2006) we can model these as emission by a
thermal plasma ({\sc apec}: Smith et al. 2001). Such models are
sensitive to the heavy metal abundance, however this is uncertain in these
sources; whilst observations of the NGC~5408~X-1 environment
(Mendes de Oliveria et al. 2006) implies sub-solar abundance, X-ray spectral fitting (Winter et al. 2007) measures abundances closer
to solar (there are no equivalent constraints for NGC~6946~X-1
although the galactic metallicity is $\sim$ solar:
Moustakas et al 2010). To account for these uncertainties we tested
abundances set at 0.5~Z$_{\odot}$ and Z$_{\odot}$. In order to be fully consistent we
included a second, variable abundance absorption column ({\sc tbvarabs}) where
we fixed the heavy metals (above He) to the abundance in the {\sc apec}
component, redshift to zero and the column density of the {\sc tbabs} component to
its Galactic value. The model fitting (Table 2) gives thermal plasma luminosities of 3.9-5.3 (NGC~5408~X-1) and 2.8-5.6$\times$10$^{38}$~erg~s$^{-1}$ (NGC~6946~X-1) for Z$_{\odot}$ and 0.5~Z$_{\odot}$; far greater than the inferred diffuse luminosity of either galaxy (Dale et al. 2005; Mineo et al. 2012) given the ratio of extraction region area to the projected galaxy area: $\sim$1$\times$10$^{37}$erg s$^{-1}$.

\begin{figure*}
\begin{center}
\begin{tabular}{l}
  \epsfxsize=15cm \epsfbox{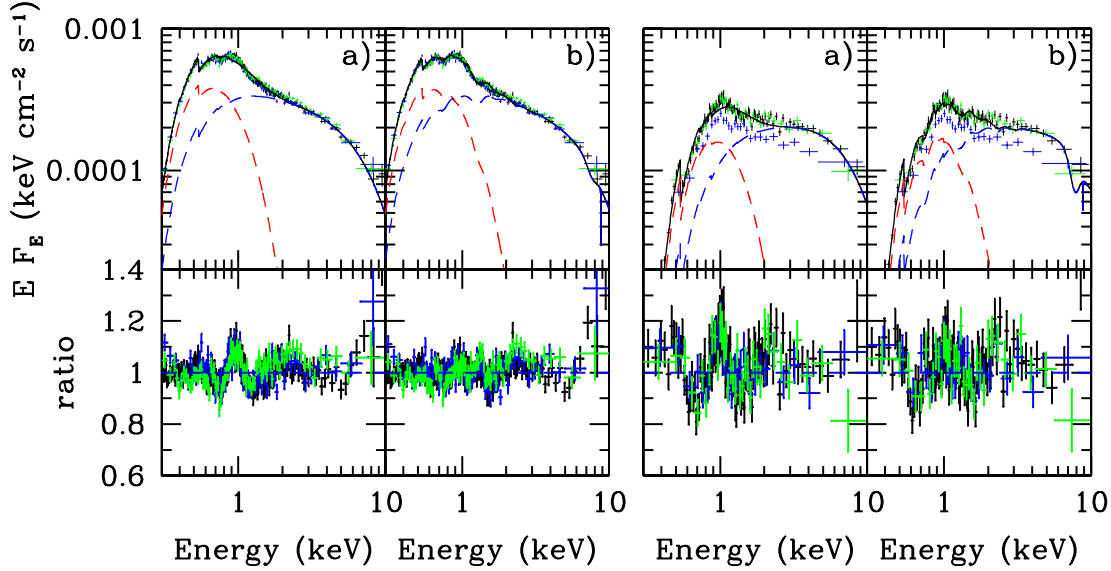}
\end{tabular}
\end{center}
\vspace{-0.4cm}
\caption{{\it Left a):} Best-fitting continuum model ({\sc diskbb}:
  red dashed line, {\sc nthcomp}: blue dashed line) to the combined
  data for NGC~5408~X-1 (PN: black, MOS1: blue, MOS2: green) with
  model parameters given in Table 2. Residuals to the fit are
  shown in the panel below and clearly show an excess at 1~keV. {\it b):} Best-fitting model incorporating a partially
  ionised, optically thin wind (see Table 2 for best-fitting
  parameters), in this case with 0.5~Z$_{\odot}$. The residuals have been
  substantially reduced. {\it Right:} as for {\it left} with the data
  of NGC~6946~X-1 and best-fitting models. Note that the MOS1 data is
  lower due to the source region being affected by a dead column.}
  \vspace{-0.3cm}
\end{figure*}

Should the residuals instead be the result of absorption in the
line-of-sight by an outflowing, optically thin phase of a
super-critical wind then by making some basic assumptions we can model
these features using a table made with {\sc
  xstar2xspec} (Kallman \& Bautista 2001). Based
on the simulations of Takeuchi et al. (2013) we may expect wind
element sizes of $\sim$10~R$_{\rm s}$. If we (naively) assume that the
column density of winds in BHBs when extremely bright ($\sim$10$^{24}$
cm$^{-2}$, e.g. Revnivtsev et al. 2002) is of order what we may expect
to see in ULXs then this places an upper limit on the particle density
of 10$^{17}$cm$^{-3}$. As this upper limit gives a Thompson
scattering optical depth $\le$1 at these columns, we are satisfied
that this is an appropriate, albeit crude value to use in this initial
modelling. We used the best-fitting de-absorbed model to the continuum to provide the input,
ionising flux, and produced four grids per source,
stepping between log$\xi$ of 3 to 4 in five linear steps and column
density of 1$\times$10$^{22}$ - 1$\times$10$^{24}$ (limited so that
the optical depth is less than unity) in eight logarithmic steps. As we
do not know the velocity dispersion {\it a priori} we set the
turbulent velocity (as a proxy for velocity dispersion) in the wind to be 0.01c and 0.1c (the predicted
outflow velocity is an obvious upper limit: Takeuchi et al. 2013) and
produced grids at 0.5~Z$_{\odot}$ and Z$_{\odot}$. In all grids we set the covering
fraction to be unity as a limiting case. Fitting the resulting grids with the continuum (again using {\sc
  tbvarabs} with matching abundances) in {\sc xspec} we find the larger velocity dispersion of
0.1c to be statistically favoured (by $\Delta\chi^{2}>50$). Whilst including smeared absorption yields a highly significant
improvement in $\Delta\chi^{2}$ from the continuum model (Table 2), this is evidently not as great as that obtained by including a thermal plasma (in both cases the fit quality remains poor for NGC~5408~X-1).  However, we  attribute this (and the spectral residuals in Figure 1) to the simplicity of our model; it only accounts for a single phase of the absorber (we do not fit multiple grids to avoid compounding caveats) yet there are likely to be multiple phases of differing densities (Takeuchi et al. 2013).

At
the relatively high ionisation states inferred for the wind, H and He
have no effective opacity and do not affect the shape of the
spectrum. As a result, the different abundances in the grids are
somewhat balanced by column density and ionisation state. Although we cannot
be certain of the correct abundance at this time, we can use both
best-fitting models (the parameters of which are given in Table 2) to
determine the likely range of parameters for the wind. Given the particle density and range of log$\xi$ we can obtain a rough
estimate for the location of the material relative to the central
illuminating source. This explicitly assumes that the wind is exposed
to the same de-absorbed, 0.3-10~keV X-ray luminosities that we observe: 7.7 and
7.6 $\times$10$^{39}$ erg s$^{-1}$ for NGC~5408~X-1 and NGC~6946~X-1
respectively (determined by integrating below the continuum model in
Table 2). Following Tarter et al. (1969): $R \ge
\sqrt{L_{\rm x}/(\xi n_{\rm e})}$, we find $R >$ 5200~R$_{\rm
  g}$ and $>$ 3400~R$_{\rm g}$ for a 10~M$_{\odot}$ BH and for each
source respectively (with the aforementioned caveats).

Using eqn 1 of Ponti et al. (2012) we can obtain a crude estimate
for the mass outflow rate in this phase of the wind:

\vspace{-0.5cm}
\begin{equation}
\dot{M}_{w} = 4\pi m_{p} v_{out}\frac{L_{x}}{\xi}\frac{\Omega}{4\pi}
\end{equation}

\vspace{-0.2cm}
\noindent where $\Omega$ is the covering fraction of the wind, $m_{p}$ is the proton mass and $v_{out}$ is the outflow velocity. We can also estimate the mass inflow rate through the inner radius using the relations of Poutanen et al. (2007): 
\vspace{-0.2cm}
\begin{equation}
\dot{M}_{o} = M_{BH}^{-1/2}\left(\frac{1.5f_{c}}{T_{c,sph}[{\rm keV}]}\right)^{2}
\end{equation}
 
\vspace{-0.4cm}
\begin{equation}
\dot{M}_{in} \approx \frac{\dot{M}_{o}(1-a)}{1 -a\left(\frac{2}{5}\dot{M}_{o}\right)^{-1/2}}
\end{equation}

\vspace{-0.3cm}
\noindent where $\dot{M}_{o}$ is the local mass accretion rate (in
units of Eddington), $T_{\rm c,sph}$ is the temperature at the wind spherization radius, $R_{\rm
  sph}$, $f_{\rm c}$ (the colour temperature correction due to Compton scattering) is assumed to be 1.7 (Poutanen et al. 2007), $a = \epsilon_{w}(0.83 -
0.25\epsilon_{w})$ and $\epsilon_{w}$ is the fraction of radiative
energy used to launch the wind. From the spectral fits we obtain
$T_{\rm c,sph}$ (kT$_{\rm d}$). If we then assume the unphysical but
limiting value of $\epsilon_{w}$ = 1 (assuming maximal energy launches the wind 
and thus largest mass loss) and use the results of the model fitting, we obtain an upper limit on the mass
outflow to inflow (to the BH) ratio of $<$ 1.1 and $<$ 0.7 for
NGC~5408~X-1 and NGC~6946~X-1 respectively. In deriving the ratio we
have assumed an accretion efficiency of 0.08 (standard for a
Schwarzchild BH) with the values differing mostly due to the lower
limit on log$\xi$ for NGC~5408~X-1. There are naturally several 
caveats e.g. this only accounts for this partially
ionised phase of the wind, and the covering fraction was set at unity (yet Walton et al. 2013 rule this out). We will explore these caveats in detail in a future work.

\section{Discussion \& conclusions}



Given the bright nature of ULXs, it is natural to assign their
properties to an exotic accretion scenario, either the presence of
IMBHs or super-critical accretion modes onto stellar mass
BHs. Assuming a common scaling of BH accretion physics (spectrum,
variability, outflows etc.), the predictions for IMBHs are clear. At
the inferred Eddington ratios we would expect generally hard spectra
(rolling over well out of the {\it XMM-Newton} bandpass), with large
amounts of fractional variability (e.g. Munoz-Darias et
al. 2011). Whilst the amount of `fast' variability in some ULXs seems
to be an adequate match (though we caution that the QPOs do not yet
provide a strong lever arm for identification), e.g. Caballero-Garcia et al. 2013, the spectra and behaviour seem to
depart from expectation (see Gladstone et al. 2009; Sutton et
al. 2013; Bachetti et al. 2013; Middleton et al. in prep). One thing is clear: at these accretion rates we do not
expect IMBHs to be powering substantial winds (assuming scaling of
accretion properties holds at these intermediate mass
ranges). Conversely if we infer the presence of accretion onto stellar
mass BHs we {\bf must} invoke super-critical accretion and winds if we
are to explain the bright ($3<$ L$_{\rm X}$ $<$ 30 $\times$10$^{39}$
erg s$^{-1}$) ULXs and their coupled spectra and 
variability (Sutton et al. 2013).

We can explain the large amounts of variability seen in some soft ULXs should they be `wind-dominated', i.e. viewed at moderate inclinations,
into the launching cone of the wind expected to accompany super-critical accretion (Middleton et al. 2011). In the cases of the two best observed of these, NGC~5408~X-1 and NGC~6946~X-1, strong residuals are seen at soft energies. Although these can be described by thermal plasma
emission, the luminosities appear improbably large for host galactic diffuse emission at the position of the source. If we instead associate these with
collisionally excited circum-equatorial material (perhaps from earlier wind epochs) then we could generate the observed
luminosities (Roberts et al. 2004). However, such emission should be isotropic, yet we appear to see these
residuals predominantly in sources with soft spectra
(and generally large amounts of fractional variability at high
energies), e.g. the `ULX dipper', NGC~55~ULX1 (Stobbart, Roberts \&
Warwick 2004).  Conversely we note the overall lack of strong
variability and residuals in those brightest and hardest ULXs thought
to be viewed face-on (Sutton et al. 2013) which are also constrained to show only very weak (or no) narrow iron features (Walton et al. 2012; 2013). This could be a remarkable coincidence or
else line-of-sight/inclination effects are an obvious solution (the lack of narrow emission in the hardest sources will be addressed in future work). 


Our analysis has shown that absorption by an optically thin plasma, outflowing into the
line-of-sight and ionised by the central source can broadly account for the
residuals. We can reconcile this within the model of super-critical
accretion where, at larger distances, the initially optically thick material in the wind disperses in the direction of the outflow (as the wind elements
themselves are not in hydrostatic equilibrium) and
becomes optically thin. Although the density of material drops,
should the combination of distance from the ionising source and
self-shielding (by the inner wind) allow the ionisation of the material
to drop, we should see the effect of absorption by ionised species of abundant elements. Due to the
velocity dispersion in the outflow, we would expect the lines to appear
smeared and blue shifted when looking into the out-flowing path of the
wind. In this way they are effectively analogous to the lines seen in
BALQSOs at lower energies. Admittedly the model we have used is only simple (as can be seen from the moderately poor fit statistics) and lacks an accurate
physical description of the outflow, radiative transfer
and effect of self-shielding. However, this first step, whilst
tentative, is still enlightening as it allows us to place initial
constraints on certain physical parameters of the wind (which will
be reviewed and updated as the models to describe the spectral imprint
of these winds develop). Based on the results of our modeling we
suggest that the optically thin wind is $>$1000~R$_{\rm g}$ from the
ionising source and may carry a large fraction of matter and energy from
the accretion flow. We argue that this identification is the most likely and is the natural expectation of super-critical accretion. However, while this interpretation is the natural expectation for the super-critical regime, we are unlikely to be able to conclusively distinguish between this and other plausible explanations until the advent of missions carrying high throughput, high energy resolution instrumentation, e.g. Astro-H (Takahashi et al. 2010).


\section{Acknowledgements}
The authors thank Tim Kallmann for insights into the use of {\sc
  xstar2xspec}. MJM acknowledges support via a Marie Curie FP7
Postdoctoral scholarship. TPR was funded as part of the
STFC consolidated grant ST/K000861/1. This work is based on observations obtained
with {\it XMM-Newton}, an ESA science mission with instruments and
contributions directly funded by ESA Member States and NASA. We thank the anonymous referee for useful suggestions.

\label{lastpage}

\end{document}